\documentclass[twocolumn]{aa}
\usepackage{graphicx}
\def\com{COM~J1740$-$5340}
\def\psr{PSR~J1740$-$5340}
\def\teff{T$_{\rm eff}$}
\begin{document}
 \title{The chemical composition of the peculiar companion to the millisecond
   pulsar  in NGC~6397\thanks{Based on observations collected  at the European
   Southern Observatory, Chile, under proposal number  69.D-0264}}


   \author{E. Sabbi\inst{1},
	 R.G. Gratton\inst{2},
	 A. Bragaglia\inst{3},
	 F.~R.~Ferraro\inst{1},
	 A. Possenti\inst{4},
	 F. Camilo\inst{5},
	 \and
	 N. D'Amico\inst{4,6}
         }

   \offprints{E. Sabbi}

   \institute{Dipartimento di Astronomia Universit\`a 
            di Bologna, via Ranzani 1, I--40127 Bologna, Italy \\
           \email{sabbi@bo.astro.it}
	\and
	   INAF--Osservatorio Astronomico di Padova,
           vicolo dell'Osservatorio 5, I--35122 Padova, Italy 
        \and
	   INAF--Osservatorio Astronomico di Bologna, via Ranzani 1,
            I--40126 Bologna, Italy 
        \and
     INAF--Osservatorio Astronomico di Cagliari,
      Loc. Poggio dei Pini, Strada 54, I--09012 Capoterra, Italy 
    \and
   Columbia Astrophysics Laboratory, Columbia
         University, 550 West 120th Street, New York, NY 10027, USA
	\and 
    Dipartimento di Fisica Universit\`a di Cagliari,
     Cittadella Universitaria, I--09042 Monserrato, Italy
     }

   \date{}

   \abstract{We present the chemical composition of the bright companion
   to the millisecond pulsar J1740$-$5340 in NGC~6397, based on high
   resolution spectra. Though the large rotation velocity of the star
   broadens the lines and complicates the analysis, the derived abundances
   are found fully compatible with those of normal unperturbed stars
   in NGC~6397, with the exception of a few elements (Li, Ca, and C).
   The lack of C suggests that the star has been peeled down to regions
   where incomplete CNO burning occurs, favouring a scenario where the
   companion is a turn-off star which has lost most of its mass. In
   addition we found an unexpected large Li abundance, which suggests
   that fresh $^7$Li has been produced on the stellar surface.

   }

 \authorrunning{Sabbi et al.}
 \titlerunning{Composition of COM J1740$-$5340}
   \maketitle
%

\section{Introduction}

\psr\ is a binary eclipsing millisecond pulsar (MSP) discovered by D'Amico et
al. (2001a, 2001b) in the globular cluster NGC~6397. A bright and variable
star, with anomalous red colours (hereafter \com), was identified by Ferraro et
al. (2001) as the companion to the MSP.  This is the first observed example of
a MSP companion whose light curve is dominated by ellipsoidal variations,
suggestive of a tidally distorted star, which almost completely fills (and is
still overflowing) its Roche lobe.

Binary evolution calculations (e.g. Tauris \& Savonije 1999;
Podsiadlowski, Rappaport \& Pfahl 2002) and the optical detections
of a few sources in the Galactic field (e.g. Hansen \& Phinney 1998;
Stappers et al. 2001) show that the most common companion to a binary MSP
is either a white dwarf or a very light (0.01--0.03 M$_{\odot}$) almost
exhausted not fully degenerate star.  In the crowded stellar environment
of a globular cluster, other kinds of companion are also possible,
resulting from dynamical encounters in the cluster core. For example,
the MSP 47 Tuc-W in 47 Tucanae is orbited by a companion whose position
in the colour-magnitude diagram (CMD) suggests that it is a main sequence
(MS) star heated by the pulsar radiation flux (Edmonds et al. 2002).

None of these hypotheses fit with the observed features of \com: it is too
luminous to be a white dwarf (V$\sim$16.6, comparable to the turn--off stars of
NGC~6397, Taylor et al. 2001, Ferraro et al. 2001); its mass ( $\sim 0.3$
M$_{\odot}$, Ferraro et al. 2003, hereafter Paper~I, Kaluzny et al. 2003) is
too high for a light stellar companion; finally, if originally it was a MS
star, its anomalous red colour would imply a bloating of the atmospheric
regions much larger than predicted by any available model (e.g. Podsiadlowski
1991; D'Antona et al. 1996). As a consequence, many intriguing scenarios were
proposed in order to explain the nature of this binary system (see Possenti
2002; Orosz \& van Kerwijk 2002; Grindlay et al. 2002 for reviews).

The unusual brightness of \com\ allows in principle a detailed study of its
chemical composition and derivation of the structural parameters of this binary
system, opening new possibilities in studying the origin of MSPs in clusters. 
In this framework we started a spectroscopic campaign with ESO telescopes; this
paper is the third in a series which reports on these observations.  In Paper~I
we presented the first results based on the high resolution spectra: we
provided the radial velocity curve, the mass ratio, the determination of the
component masses, a preliminary evaluation of the metallicity and a discussion
of the heating of the companion surface due to the impinging MSP flux. In Sabbi
et al. (2003, Paper~II) we showed the complex structure of the H$\alpha$ line
(deriving important information on the mass loss) and discussed the unexpected
presence of He {\sc i} lines, implying the existence of a very narrow heated
region on the companion surface. This paper focuses on the chemical abundance
analysis.  The observational data are presented in \S 2, while in \S 3 we
report on the rotational velocity, the atmospheric parameters and the
equivalent widths of the observed spectral lines. \S 4 is devoted to the
determination of the actual element abundances, which are discussed in \S 5.

\section{Observations and data reduction}

Eight high resolution spectra of \com, taken at different phases and covering
the whole orbital period of the binary system ($P\sim$1.35 days) were acquired
with the {\it Ultraviolet-Visual Echelle Spectrograph} (UVES) mounted at the
{\it Kueyen} 8m-telescope (UT2) of the ESO Very Large Telescope on Cerro
Paranal (Chile). More details on the observational strategy can be found in
Paper~I.  \com\ is located in a crowded region, hence seeing conditions are
relevant to minimize contamination from nearby objects: all the spectra (but
one) were taken with seeing between 0.9 \arcsec and 1.2 \arcsec. Our spectra
cover, with some gaps, the wavelength range 3280--6725 \AA, at a resolution R
$\sim$ 40000; the S/N of the individual spectra varies, but it is typically
about 30--35 near H$\alpha$.

In the following analysis we started with the one-dimensional, wavelength
calibrated spectra produced by the UVES pipeline (Ballester et al. 2000). All
our abundance analysis is applied to the sum of two spectra acquired near the
orbital quadratures (at orbital phases $\sim$ 0.02, and 0.56 respectively),
when the radial velocities are highest and the larger Doppler shifts ensure
that contamination of lines by the sky and/or by other cluster stars is
negligible (Paper~I). After excision of cosmic rays and shift to zero radial
velocities, we combined the spectra and rebinned them with 0.3 \AA\ resolution
in order to enhance the S/N ($\sim 60$ near H$\alpha$) up to the level required
by the purposes of the present paper; there is no loss of information in doing
so, given the width of the lines.

\section{Analysis and error estimates}

\subsection{Rotation Velocity}  

The lines of \com\ are greatly broadened by the rotation velocity of the
star. In order to evaluate this quantity, we exploited the spectra of
three subgiants (SGB) in NGC~6397 (i.e. stars with similar temperature and
gravity). They were observed using a UVES configuration similar to ours
within a project devoted to study the chemical composition of NGC~6397
(Gratton et al. 2001, hereafter G01).  The positions of the three
aforementioned subgiants and \com\ are highlighted in the CMD shown
in Fig.~1.

In order to derive the rotation velocity we selected the subgiant
observed with the highest S/N ratio, namely \#~793 in G01
([Fe/H]\footnote{We adopt the usual spectroscopic notation that
[A/B]$=log_{10}(N_A/N_B)_{star}-log_{10}(N_A/N_B)_{\odot}$ for element A
and B}=$-2.04$, T$_{\rm eff}$= 5478\,K, $\log g$=3.4, S/N $\sim$ 100).
It represents a very good template for a "not rotating" star: a firm
upper limit to the projected rotation velocity has been recently set at
V$\sin i \leq$ 3.7 km s$^{-1}$ (Lucatello \& Gratton 2003).

A cross-correlation technique was applied to compare the spectrum of the
selected subgiant with that of \com.  
 This technique gives an upper limit to the real rotation velocity, because
the cross correlation peak is widened by factors other than rotation (line
width of the template star; micro- and macro-turbulence for the program star;
instrumental profile). However, in our case these other widening factors are
much smaller than those due to rotation of \com: the FWHM of the profiles of
the template lines is 10.5~kms$^{-1}$; typical macro-turbulent velocity for
subgiants of NGC~6397 is 6.7~kms$^{_1}$ (Lucatello \& Gratton 2003); the
microturbulent velocity for \com\ is about 1.0 km s$^{-1}$ (see below); and
finally, the instrumental broadening is about 6 kms$^{-1}$. When we sum in
quadrature all these terms, the total contribution to line width is 13.9 km/s.
We can then safely assume that the rotational velocity of \com\ (expected to be
about 50 kms$^{-1}$) dominates the width of the cross-correlation peak.

As the blue part of the spectra is richer in lines and lacks telluric features,
three different portions of this spectral region (namely $\lambda\lambda$
4120--4180, 4120--4310, and 4361--4441 \AA) were selected, avoiding strong
lines like Ca {\sc ii} H and K, and Balmer. We have determined the FWHM (i.e.
the rotational velocity) of the three cross-correlation peaks, fitting them
with a Dirac $\delta$ function broadened by rotation effects, assuming a gray
atmosphere, a normal limb-darkening law (coefficient equal to 0.6, appropriate
for stars of this temperature), and several velocities. The $V \sin i$, where
$i$ is the orbit inclination, in the three spectral regions are (see
Fig.~\ref{vrot}) 51.0 $\pm$ 1.5, 49.0 $\pm$ 1.5, and 48.0 $\pm$ 2.1 kms$^{-1}$
respectively ($1\sigma$ errors); the weighted average is $V \sin i$ = 49.6
$\pm$ 0.9 kms$^{-1}$. A fully compatible value  ($V \sin i=48.6\pm 0.9$
kms$^{-1}$ has been obtained by subtracting quadratically the broadening terms
due to the other factors mentioned above.

\begin{figure}
\centering
\includegraphics[width=8.7cm]{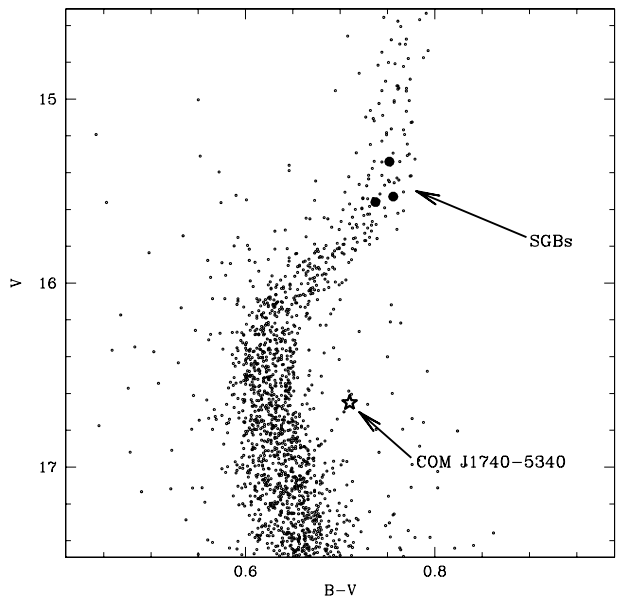} 
\caption{Position in the colour-magnitude diagram of the three subgiant
stars examined in G01 (filled dots) and of \com\ (star). \com\ B and V
magnitudes are taken from Kaluzny et al. (2003, Table 1, in quadrature).}
\label{CMD}
\end{figure}

\begin{figure}
\centering
\includegraphics[width=8.7cm]{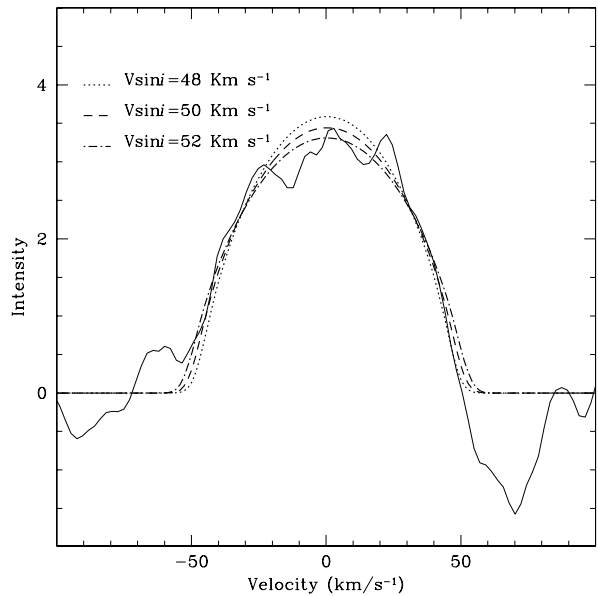} 
\caption{Peak of the cross correlation between \com\ and star \# 793 of G01
(solid line) in the spectral region included between $\lambda\lambda$ 4120--4180
\AA, and examples of three possible choices of V$\sin i$ (broken
lines): the best fit in this case turns out to be 51 km s$^{-1}$.}
\label{vrot}
\end{figure}

\begin{figure}
\centering
\includegraphics[width=8.7cm]{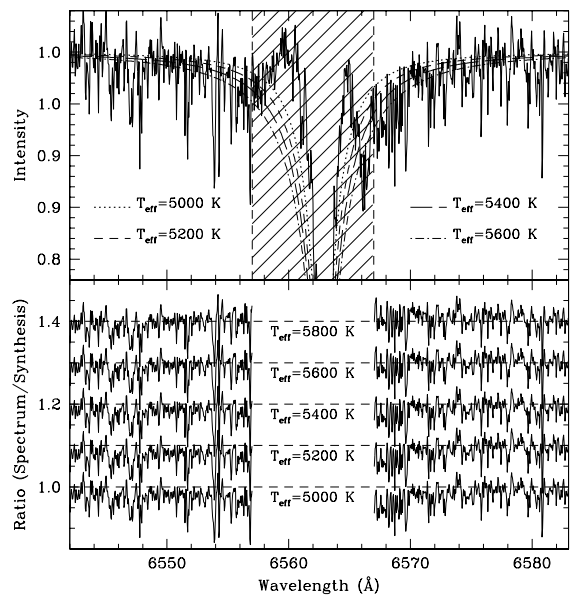}
\caption{Determination of effective temperature from the H$\alpha$
wings.  The upper panel shows the observed H$\alpha$ spectrum and a set
of synthetic Kurucz models at different \teff's; the shaded region is
excluded from the fit.  The lower panel shows the corresponding ratios:
the dashed horizontal lines represent ratio = 1 for each case; note that
the best fit cases are for \teff = 5400 and 5600 K.}
\label{temp}
\end{figure}

\subsection{Atmospheric parameters: temperature and gravity}


The \com\ effective temperature (\teff) has been determined using two different
approaches: examining the H$\alpha$ lines in our spectra (a reddening
free method, see e.g., G01) or adopting the broad band colours
resulting from the photometric observations of Kaluzny, Rucinski \&
Thompson (2003).  Unfortunately, too few lines were available to
derive the \teff ~from line excitation.

\vspace{0.3cm}
\noindent $\bullet$ The shape of the H$\alpha$ absorption line wings is a
good temperature indicator for \teff ~higher than about 5000 K (Fuhrmann
et al.  1993); the core is best left out in all stars, and even more in our case,
where an emission component is present (Paper~II).  Although this method
assumes a ``normal'' atmosphere for \com, which is not strictly true
(e.g., the star is not spherical), it is likely a good approximation
for the deep regions where the H$\alpha$ wings forms. Even if there
are other Balmer lines in our spectrum, we worked only on H$\alpha$
because it is less contaminated by metal lines, and less dependent on
metallicity, gravity, and details in the convection treatment.

As a first evaluation, we tried to reproduce the H$\alpha$ wings with
synthetic spectra based on the appropriate Kurucz (1995) model
atmosphere, without overshooting: we assumed [Fe/H]=$-2$ (G01), and
log $g$ = 3.5. Several different T$_{\rm eff}$'s have been adopted (the
cases at 5000, 5200, 5400, and 5600 K are shown in Fig.~\ref{temp},
top panel), and the best fit temperature appears to be \teff = 5500
$\pm$ 100 K. The best fit temperature has also been determined by the
ratio of the observed and synthetic profiles (Fig.~\ref{temp}, lower
panel, where identity between the two profiles is indicated by the
horizontal lines) yielding the same \teff.  In these estimates, the
adoption of a metallicity is the most important factor, while the
surface gravity impact on \teff~ is small, (an error of 0.3 
dex on log $g$ involves an error of 54 K on the evaluation of T$_{\rm eff}$).

\noindent $\bullet$ The second method uses the photometric colours and
implies independent knowledge of the reddening.  Since the star is
variable, we chose the average of the values published in the Table
1 of Kaluzny, Rucinski \& Thompson (2003), obtaining (B-V)$_{avg}
\sim$0.74. We also assumed E($B-V$) = 0.183, as recently derived from
Str\"omgren and Johnson photometry (Gratton et al. 2003).  From the
appropriate Kurucz transformation, we derived \teff = 5560 $\pm$ 100 K.
We did not use $V-I$ for an independent determination, because we had
only the Johnson-Cousins colour available, while Kurucz tables require
both $V$ and $I$ in the Johnson system and this transformation is extremely
uncertain in a H$\alpha$ emitting star, like \com.

In summary, averaging the two temperatures, we obtain \teff = 5530 $\pm$
70 K, in good agreement with previous estimations (Ferraro et al. 2001;
Paper~I; Orosz \& van Kerkwijk 2003).

The surface gravity $g$ has been estimated from the position of \com\ in the
CMD and from its mass, as derived in Ferraro et al. 2001, 2003
respectively. Since the latter is constrained to be in the interval
$0.22-0.32$ M$_\odot$, we took a value of 0.3 M$_\odot$ and derived log $g$ =
3.5 $\pm$ 0.2, which is confirmed both by the following abundance analysis and
the ionization equilibrium.

\subsection{Equivalent Widths}

The equivalent widths (EWs) measured in our spectra have been used
both for refining the set of atmospheric parameters (surface
gravity from the equilibrium of ionization, and microturbulent
velocity from elimination of trends with expected line strength) and
for deriving the actual element abundances. For some elements (e.g.,
Li) and molecules (CH) we also used spectral synthesis. A general
problem encountered in the analysis is that the rotation washed out
the weak, high excitation lines, the ones most directly depending on
the element abundance. We are thus left with strong and saturated
lines, whose analysis is dependent on the microturbulent velocity
$v_{\rm t}$ and on the damping, respectively. For the latter we used
the Barklem et al. (2000) treatment, the best available at the present
time.

As a first step, EWs were measured on the unidimensional, extracted
spectral orders using a gaussian function and an automatic routine
working within the ISA package (Gratton 1988). A few lines, missed by
the automated search, were added manually. Finally, the EWs actually
used for the abundance analysis were derived from the relation between
line depth and equivalent width (for a detailed description of the
procedure see Bragaglia et al. 2001). The list of lines measured and
used in the abundance analysis is given in Table 1, together with the
adopted excitation potentials (E.P. in eV), the oscillator strengths
(log $gf$'s), the observed EW in m\AA~  and the derived abundances.
The typical error on the EWs is about 5 m\AA. Note that the measured
EWs are influenced by rotation since the line broadening often produces
blends. A discussion of the line broadening relevance for determination
of the elemental abundance is given in next section.

We measured lines both of Fe {\sc i} and Fe {\sc ii}; from the equilibrium of
ionization we obtained a surface gravity value ($\log g = 3.46 ~\pm 0.2$) in
perfect agreement with the one derived from photometry.

Optimization of the microturbulent velocity was done by comparing the
abundances given by the intermediate strength lines with those derived
from the strong ones, with well developed damping wings. In fact this
would lead to a $v_{\rm t}$ quite smaller than required to get a good
ionization equilibrium, so we compromised on an average value,
adopting $v_{\rm t}$ = 1 ($\pm$ 0.5) kms$^{-1}$.

In summary, the abundance analysis has been performed adopting the
following atmospheric parameters: \teff = 5530 K, log $g$ = 3.46, $v_{\rm
t}$ = 1.0 kms$^{-1}$; by the analysis of 12 Fe lines, we derive [Fe/H] =
$-1.85 \pm 0.13$ (see next Section for a better assessment of the errors).
These values are to be compared with those used by G01 in the analysis
of the NGC 6397 subgiants: \teff = 5478 K, log $g$ = 3.42, $v_{\rm t}$
= 1.32 kms$^{-1}$ and [Fe/H] = $-2.05 \pm 0.03$.

\begin{table}
\begin{center}
\label{t:abu1}
\caption[]{Equivalent widths (only for the most reliable lines) measured
on the \com\ spectrum and derived abundances. Li {\sc i}, Ti {\sc ii},
and Sr {\sc ii} abundances come from synthetic spectra (SS flag in
the last column), while Na {\sc i} abundances have been corrected for
departure from local thermodinamical equilibrium (NLTE).}
\begin{tabular}{lccrrcl}
\hline\hline
\\
Elem.    & $\lambda$ & E.P. & log $gf$ & EW   & abund.  & Notes \\
	 &   (\AA)   & (eV) & 	       & (m\AA) &  & 	\\
\\
\hline
\\
Li {\sc i}  & 6707.80 	     & 0.00 &  0.19    &       & 2.2  & SS\\
Na {\sc i}  & 5889.97 	     & 0.00 &  0.18    & 167.2 & 4.31 & NLTE\\
Na {\sc i}  & 5895.94 	     & 0.00 & -0.12    & 167.2 & 4.61 & NLTE\\
Mg {\sc i}  & 5172.70 	     & 2.72 & -0.32    & 247.1 & 5.66 & \\
Mg {\sc i}  & 5183.62 	     & 2.72 & -0.10    & 278.3 & 5.56 &\\
Mg {\sc i}  & 5528.42 	     & 4.34 & -0.52    &  64.0 & 5.64 &\\
Ca {\sc i}  & 6122.23 	     & 1.89 & -0.27    &  82.0 & 5.07 &\\
Ca {\sc i}  & 6439.08 	     & 2.52 &  0.39    &  71.5 & 4.81 &\\
Ti {\sc ii} & 4395.03        & 1.08 & -0.51    &       & 3.4  & SS\\
Fe {\sc i}  & 4045.81 	     & 1.49 &  0.28    & 317.2 & 5.79 &\\
Fe {\sc i}  & 4063.59 	     & 1.56 &  0.06    & 267.1 & 5.91 &\\
Fe {\sc i}  & 4071.74 	     & 1.61 & -0.02    & 227.9 & 5.87 &\\
Fe {\sc i}  & 4202.04 	     & 1.49 & -0.71    & 124.7 & 5.74 &\\
Fe {\sc i}  & 4383.56 	     & 1.49 &  0.20    & 275.7 & 5.68 &\\
Fe {\sc i}  & 4404.76 	     & 1.56 & -0.14    & 192.9 & 5.73 &\\
Fe {\sc i}  & 5325.19 	     & 3.21 & -0.10    &  85.0 & 5.84 &\\
Fe {\sc i}  & 5405.78 	     & 0.99 & -1.84    &  82.5 & 5.48 &\\
Fe {\sc i}  & 5434.53 	     & 1.01 & -2.12    &  62.5 & 5.29 &\\
Fe {\sc ii} & 4923.93 	     & 2.89 & -1.35    &  91.8 & 5.89 &\\
Fe {\sc ii} & 5018.45 	     & 2.89 & -1.22    &  95.4 & 5.84 &\\
Fe {\sc ii} & 5316.62 	     & 3.15 & -2.02    &  52.7 & 5.81 &\\
Sr {\sc ii} & 4077.71        & 0.00 &  0.17    &       & 1.1  & SS\\
Ba {\sc ii} & 6141.75 	     & 0.70 &  0.00    &  60.8 & 0.62 &\\
\\
\hline
\end{tabular}
\end{center}
\end{table}

\begin{table}
\begin{center}
\label{t:abu2}
\caption[]{Abundances of \com\ (in column 2), compared to those of
normal SGBs; see G01 for Fe, Na, and Mg, while the other values are
still unpublished (see text).}
\begin{tabular}{lccc}
\hline\hline
\\
Element	& abund. & abund. &  $\Delta$(abund.)\\
      & COM    & SGB &     \\
\\
\hline \\
Fe {\sc i}   & 5.70   &   5.42  & +0.28  \\
Fe {\sc ii}  & 5.84   &   5.26  & +0.58  \\
Li {\sc i}   & 2.2    &   1.2   & +1.00  \\
Na {\sc i}   & 4.46   &   4.51  & --0.05 \\
Mg {\sc i}   & 5.62   &   5.70  & --0.08 \\
Ca {\sc i}   & 4.94   &   4.54  & +0.40  \\
Ti {\sc ii}  & 3.4    &   3.39  &  0	 \\
Sr {\sc ii}  & 1.1    &   0.75  & +0.35  \\
Ba {\sc ii}  & 0.62   &   -0.06 & +0.68  \\
\\
\hline
\end{tabular}
\end{center}
\end{table}

\begin{figure*} \centering
\includegraphics[width=16cm,height=14cm]{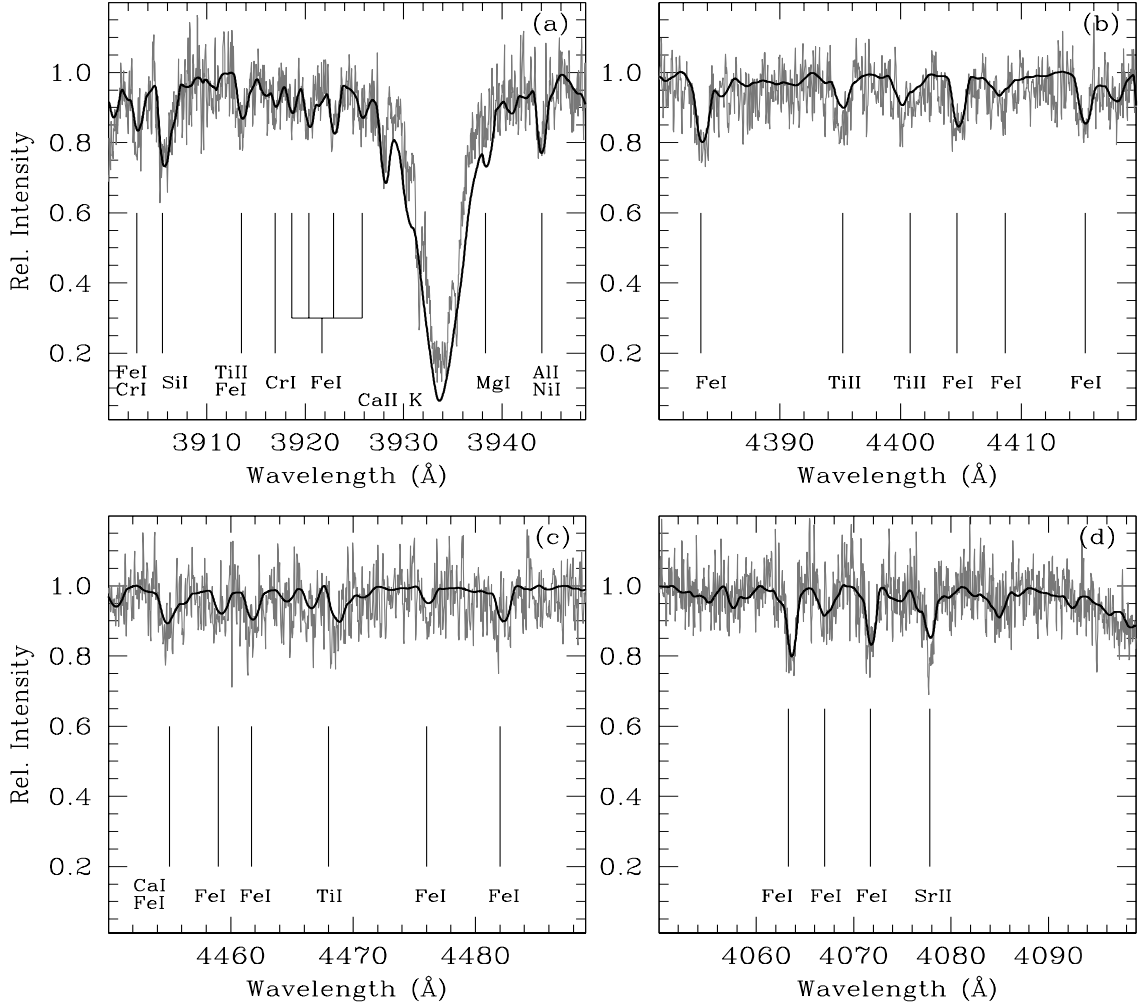}  
\caption{Comparison between the \com\ spectrum (grey line) and the
subgiants template (black line, obtained by averaging, and broadening
to account for rotation, three normal NGC 6397 subgiant stars,
G01). We show the spectral regions near lines of Ca {\sc ii} (panel
(a)), Ti {\sc ii} (panel (b)) Ti {\sc i} (panel (c)) and Sr {\sc ii}
(panel (d)). In all these regions many Fe {\sc i} lines are also
present.}
\label{comp1} 
\end{figure*}

\begin{figure*}
\centering
\includegraphics[width=16cm,height=8cm]{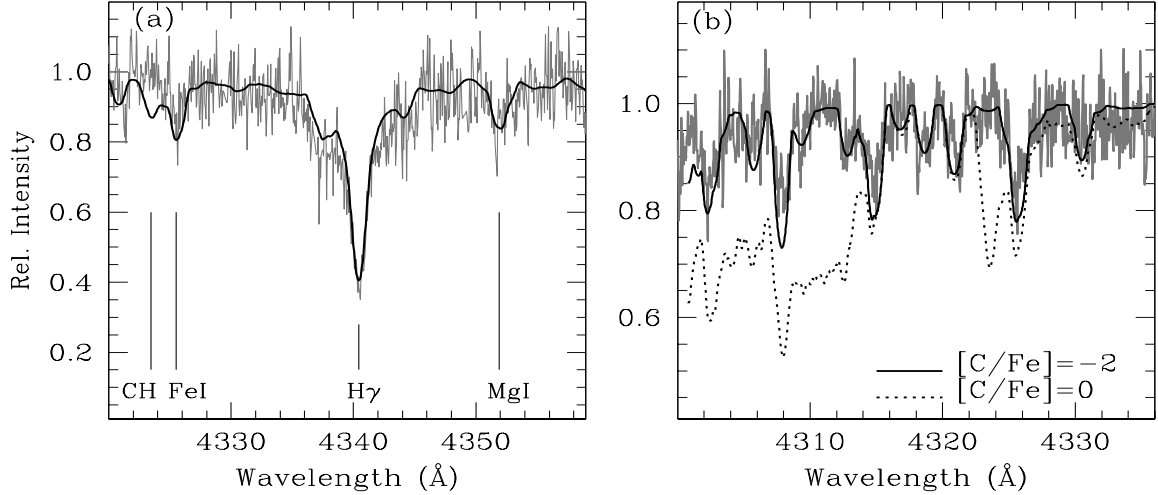}
\caption{Evaluation of the carbon abundance from the CH band. Panel
(a) shows a clear depletion in the region of the CH band in the \com\
spectrum (grey line) with respect to the NGC 6397 subgiants template
(black line). Panel (b) shows the result of spectral synthesis,
demonstrating that C is strongly underabundant in the \com\
atmosphere.}
\label{ch}
\end{figure*}

\begin{figure*}
\centering
\includegraphics[width=16cm,height=9cm]{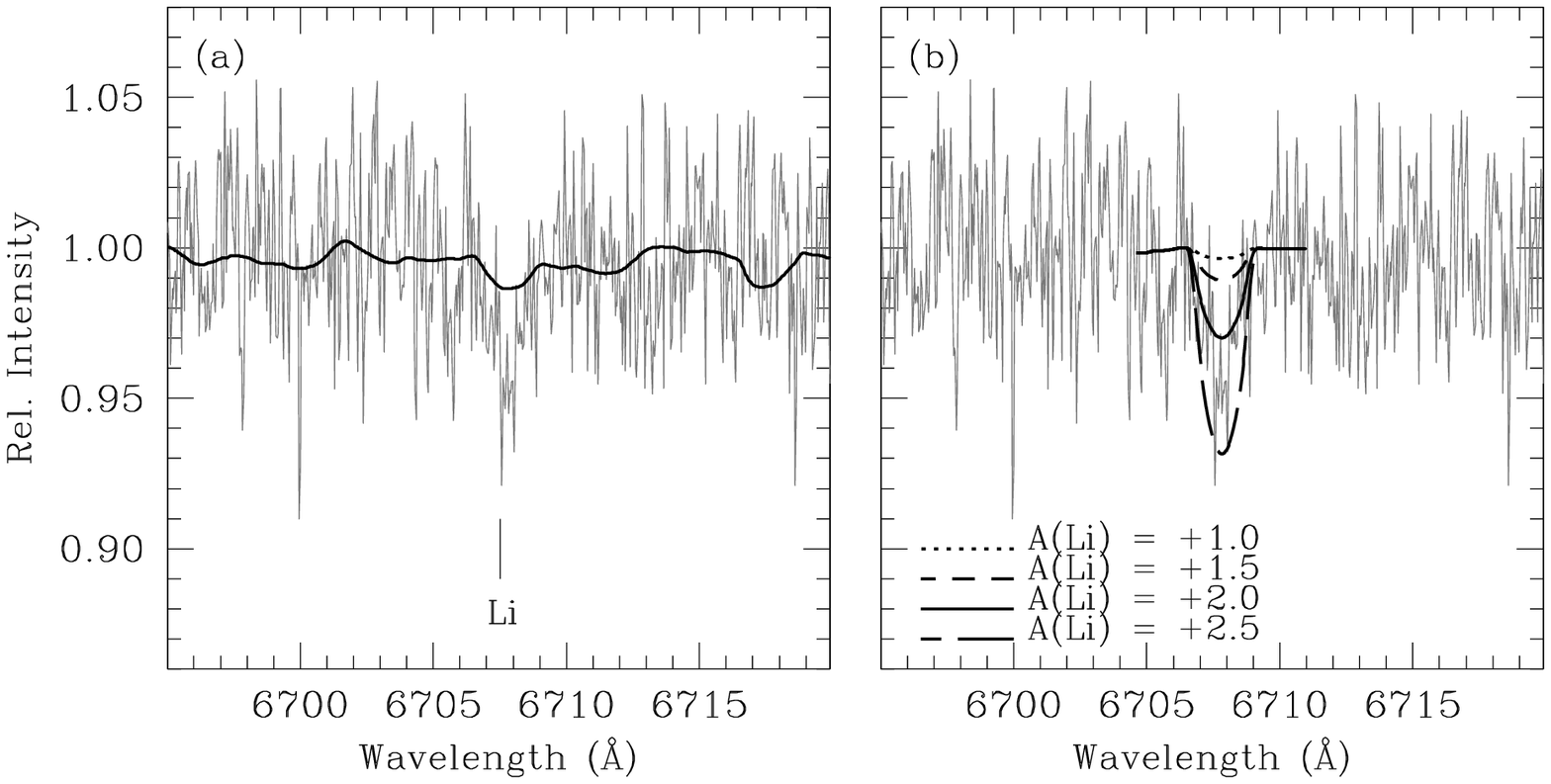}
\caption{In panel (a) we compare the \com\ spectrum with the NGC 6397
subgiants template: the Li {\sc i} line is clearly deeper than in the
template. In panel (b) we show a comparison with four different
synthetic spectra, demonstrating the high Li content.}
\label{li}
\end{figure*}

\section{The chemical abundance}

How do the element abundances in \com\ compare with those measured in
other stars of NGC 6397? To address this question, we have considered
three NGC 6397 SGB stars observed by G01 with a similar UVES
configuration, and which display very similar atmospheric parameters
(see \S 3.1). Combining the spectra of the three stars and convolving
the summed spectrum with a rotational profile of 49.6 km\,s$^{-1}$ (see
\S 3.1), we have built a high S/N template spectrum.

We can first exploit this template spectrum for assessing the typical
uncertainty in our abundance analysis due to the effects of the rotation of COM
J1740$-$5340. In fact, comparing the line EWs both in the original SGB summed
spectrum and in the spectrum broadened by rotation, we have noticed a
systematic increasing in the abundances measured from the latter (even of 0.5
dex).  This confirms that no immediate comparison should be done with the
chemical composition of non-rotating stars, since \com\ lines are broadened:
blends with nearby lines can increase the EWs, and the derived abundances may
be in error. Another very important factor is that the atmospheric model for
\com\ is not adequate, given the non spherical --- and possibly variable ---
geometry of COM J1740$-$5340: this mostly affects the ionized species, whose
abundances are less reliable. From these consideration an error of 0.2 dex on
the derived metallicity of  [Fe/H] =$-$1.85 was considered, while the other
abundances are determined with larger errors, about 0.3 dex.

Inspection of Fig.~\ref{comp1} (where they have been superimposed in four
different regions comprising lines of Ca, Ti, Mg, Sr and Fe) demonstrates
that the \com\ and the SGB template spectra show an excellent agreement,
even if the \com\ spectrum presents some evidences of lines filling (they
are evident mostly in the core of the Ca {\sc ii} K line). Since the
latter feature can be ascribed to the emission by the warmer region of
the atmosphere (also responsible for the He {\sc i} lines: see Papers~I \&
II), this comparison supports the hypothesis that the chemical composition
of \com\ is similar to that of the SGB stars.

This conclusion would nominally contrast with the values collected
in Table 2, where the forth column lists the differences between the
average values of the abundances for each species measured in \com\
and those derived by G01 for normal subgiant not rapidly rotating
stars in NGC 6397. Taking these estimates at face value, \com\ would
show an overabundance of various species, particularly iron. However,
according to the aforementioned difficulties involved in the analysis,
we conclude that, within the errors, the abundances of Fe, Na, Mg, and
Ti in \com\ are compatible with those measured in normal NGC 6397 stars.

\subsection{Discrepant elements: Ca, C, Li}

In the previous section we have shown that, once the broadening effects on EWs
are taken into account, the chemical abundances of \com\ are fairly similar to
those of SGB stars. Nevertheless, even after adjustment for rotation, some
elements still show disagreement in abundances with the other NGC 6397 stars.
Since for the Sr {\sc ii}  and Ba {\sc ii} only one line (cominng from ionized
species, whose  abundance determination is less reliable) each is present in
our spectra, we do not consider here these two elements, and  focus only on the
three others.

(i) Ca: We measured only four Ca {\sc i} lines in our spectra, among which are
the very strong ones at $\lambda\lambda$ 4226.74 \AA ~and 6162 \AA. Since the
strength of these lines depends more on $v_{\rm t}$ than on abundance (hence
they are rarely used), we retained only the two most reliable for abundance
determination (namely at $\lambda\lambda$ 6122 \AA ~and 6439 \AA). We
accurately checked our spectrum near the lines for cosmic rays and hidden
blends, and directly compared it to the SGB template. \com\ Ca {\sc i} lines
appear deeper than in the template, as a conseguence there are differences in
the EWs (e.g., about 20 m\AA\ for the 6122 \AA\ line). Taking into account the
slightly different atmospheric parameters for \com\ and the SGB template, this
implies a Ca {\sc i} abundance about 0.4 dex higher for our star.

(ii) C: Comparing our spectrum to the template SGB one, it is clear
that C is strongly underabundant in \com\ (Fig.~\ref{ch}(a)). To get a
quantitative estimate of the C abundance we have built synthetic spectra
for the region near the CH G-band (around 4310 \AA) by assuming [C/Fe] =
$-2$, $-1$, $0$, $+1$: two of them are shown in Fig.~\ref{ch}(b). The
result is that \com\ spectrum is fully compatible with complete absence
of C in the atmosphere.  This would indicate a composition resulting
from the CN cycle at equilibrium, when most C has been burned to N;
this evidence suggests that \com\ could be a deeply peeled stars.

Ergma \& Sarna (2003) computed the evolution of the surface abundances
for a few key elements in two cases proposed for explaining the nature
of \com\ (low mass bloated main sequence star or evolved star). From
their calculations C, N, and O would behave like in ``normal'' metal
poor stars if \com\ is a low mass main sequence star. On the contrary,
only O would have a normal abundance, while N would be overabundant,
and C underabundant in the case of a heavier, evolved star which has
lost mass.  Unfortunately, no O and N lines are present in our spectra,
but the very low C abundance favours the second possibility.

(iii) Li: The Li {\sc i} resonant doublet at 6708 \AA\ is much
stronger in our spectrum than in the normal NGC 6397 subgiants: Fig.
~\ref{li}(a) shows a comparison between \com\ and the SGB template,
while Fig. ~\ref{li}(b) is a zoom in on the Li feature with a few
synthetic models superimposed. From the best fit synthetic spectrum we
obtained a Li abundance (A(Li) = 2.2 ($\pm$ 0.2) in the usual notation),
which is quite surprising. In fact, the Li abundance of turn-off stars,
i.e. of unevolved stars with the original abundance, is A(Li) = 2.34
$\pm$ 0.08 (Bonifacio et al. 2002), while subgiant stars, which have
already diluted their original Li, have A(Li) $\sim$ 1.5 (Castilho et al.
2000), a value also confirmed by A(Li) $\sim$ 1.2 measured in the three
SGB stars used as template (see Table 2).

This value is at odds with what is expected both from a $\sim$ 0.3
M$_\odot$ main sequence star recently acquired in the binary system
(since low mass stars are completely convective and the original surface
Li has long been transported to the interior and there destroyed) and from
a heavier star peeled by mass loss down to the zone where the CN cycle
is active (incomplete CNO process, burning C to N).  How can we explain
this high Li content? It cannot be the original, cosmological Li, nor
Li produced by mechanisms such as the Hot Bottom Burning, which works in
intermediate mass stars (Ventura et al. 2001).  Planet engulfing
(Alexander 1967; Siess \& Livio 1999) is also improbable: the orbit of a 
planet around \com\ could not be stable and Li would not be the only element
affected (e.g., C too would be enriched, not depleted).

At the moment the most plausible explanation could be fresh Li production
due to nuclear reactions occurring on the stellar surface, since a nearby
source of cosmic rays (the pulsar) is available. In particular, it should
be noticed that a peeled star, where we see at the surface the products
of incomplete CNO cycle, should have a large atmospheric abundance of
$^3$He: $^7$Li can be easily produced by a cosmic ray flux containing
$\alpha$-particles (D'Antona, private communication). Moreover, at
the very high temperature required (a few million K), the star would
also be an X-ray source, and this is actually observed (Grindlay et
al. 2001, 2002), even if the X-ray emission can be also attributed to
the interaction between the relativistic wind from the MSP and the mass
lost from the companion.  Finally, we recall that Li enhancement, found
in some chromospherically active binary systems, is actually attributed
to nucleosynthesis in the corona, i.e., to a similar mechanism (see,
e.g. Montes et al. 1997, 2000 for a review).

\subsection{He {\sc i} at $\lambda$ 6678 \AA}

\begin{figure}
\centering
\includegraphics[width=9cm]{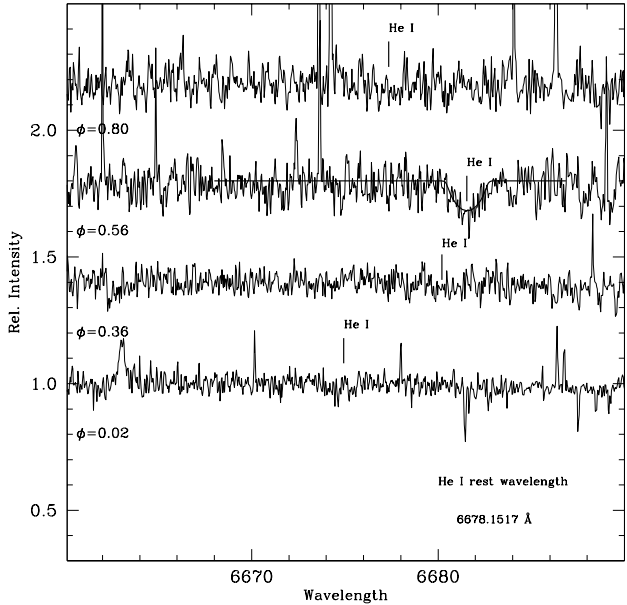}
\caption{He {\sc i} line at $\lambda$ 6678 \AA\ at different orbital
phases. The heavy solid line overlaid on the spectrum acquired at phase
$\phi$=0.56 (quadrature) is an empirical profile broadened by stellar
rotation. In each spectrum the He {\sc i} position is indicated.}
\label{He66}
\end{figure}

As discussed in Papers~I and II the inspection of the spectral data-set
reveals the unexpected presence of He {\sc i} lines at $\lambda$ 5876
\AA~ and $\lambda$ 6678 \AA.  In particular the He {\sc i} properties
at 5876 \AA~ were fully described in Paper~II. Here we present the phase
variations of the He {\sc i} line at 6678 \AA~ (see Fig.~\ref{He66}). As
can be seen from the figure, such line is clearly visible (and measurable)
only at $\phi$= 0.56 (quadrature). Note that at such a phase the
line has the same EW as the line detected at 5875 \AA\ (0.19 \AA
see Paper~II, Fig. 4). At the other orbital phases the S/N is not
good enough to allow a reliable evaluation of the line EW. At $\phi$=
0.36 (near conjunction) the line is practically absent. This in not a
surprising result since at this phase also the 5878 \AA~ line is shallower
than at the other phases. At $\phi$=0.02 (quadrature) the He line is
very shallow; moreover a cosmic ray was present just at the He {\sc i}
wavelength and, even if we removed it, it is possible that the line is
still partially contaminated by this cosmic ray.  Spectra at higher S/N
are required to clearly assess the presence of the line at this phase.

\section{Summary and Conclusions}

We have presented the chemical composition of \com, the non-degenerate
companion to the millisecond pulsar \psr\ in NGC 6397. Abundance
analysis of high resolution spectra has allowed us to determine the
abundance of iron ([Fe/H] = $-1.85 \pm 0.2$) and several other
elements. The general conclusion is that the abundance is fully
compatible with that of normal, single stars in NGC 6397, with a few
notable exceptions (Ca, C, and Li) that may be attributable to the
peculiar history of the star, subject to (extreme) mass loss and
interactions with the millisecond pulsar. In particular, the strong C
depletion seems to indicate that \com\ is not a perturbed low mass main
sequence star, but had instead a larger mass and has been peeled down
to the present $\sim 0.3$ M$_\odot$. Future observations will allow
the appropriate measure of N and O abundances, and would give further
support to this hypothesis, if those elements are found overabundant
and unchanged, respectively, with respect to other stars in NGC 6397.

From the derived chemical composition, we do not see any indication of
accretion of elements from the type II supernova explosion that left
behind the neutron star (e.g., Mg), since the composition of \com\ is
mostly similar to the other stars in the cluster. Several scenarios,
among which we are not able to discriminate with the present data,
are possible: i) the secondary has been acquired by the system
only after the SN (star exchange in a collision); ii) the SN wind
was too fast for the secondary star to accrete a significant amount of
ejected material; iii) and/or the neutron star has had enough time to
remove all the accreted material, together with a fair fraction of its own
original mass, from the companion.

\begin{acknowledgements}
We thank E. Carretta and L. Cinque for useful comments and discussions. This
research was supported by the {\it Agenzia Spaziale Italiana} (ASI).
\end{acknowledgements}

\end{document}